\documentstyle[preprint,eqsecnum,prd,aps,epsf]{revtex}   
\newcommand{\DO}{D\raise1pt\hbox{$\not$}O}
\begin{document}
\draft
\tightenlines
\preprint{FERMILAB-Pub-00/160-E}
\title{
Examination of direct-photon and pion production \\ 
in proton--nucleon collisions}

\author{
L.~Apanasevich,$^{2,3}$
M.~Begel,$^{3}$
C.~Bromberg,$^{2}$
T.~Ferbel,$^{3}$
G.~Ginther,$^{3}$
J.~Huston,$^{2}$
S.~E.~Kuhlmann,$^{1}$
P.~Slattery,$^{3}$
M.~Zieli\'{n}ski,$^{3}$
V.~Zutshi$^{3}$
}
\address{
\centerline{$^{1}$Argonne National Laboratory, Argonne, Illinois 60439}
\centerline{$^{2}$Michigan State University, East Lansing, Michigan 48824}     
\centerline{$^{3}$University of Rochester, Rochester, New York 14627}          
}
\date{\today}
\maketitle
\begin{abstract}
We present a study of inclusive direct-photon and pion production in
hadronic interactions, focusing on a comparison of the ratio of
$\gamma/\pi^0$ yields with expectations from next-to-leading order
perturbative QCD (NLO pQCD). We also examine the impact of a
phenomenological model involving $k_T$ smearing (which approximates
effects of additional soft-gluon emission) on
absolute predictions for photon and pion production
and their ratio.
\end{abstract}


\narrowtext

\section{INTRODUCTION}

Direct-photon production in hadronic collisions at high transverse
momenta ($p_T$) has long been viewed as an ideal testing ground for
the formalism of pQCD.  A reliable theoretical description of the
direct-photon process is of special importance because of its
sensitivity to the gluon distribution in a proton through the
quark--gluon scattering subprocess ($gq\rightarrow\gamma q$).  The
gluon distribution, $G(x)$, is relatively well constrained for $x<0.1$
by data on deep-inelastic lepton--nucleon scattering~(DIS) and
dilepton (``Drell-Yan'') production, and by collider data on jet
production at moderate $x$ ($0.1-0.25$)~\cite{cteq4}, but is
less constrained at larger~$x$~\cite{huston-uncertainty}.  In
principle, fixed-target direct-photon production can constrain $G(x)$
at large~$x$, and such data have therefore been incorporated in
several modern global parton distribution function (PDF)
analyses~\cite{cteq4,grv92,mrst}.

\begin{figure}
\epsfxsize=3truein
\centerline{\epsffile{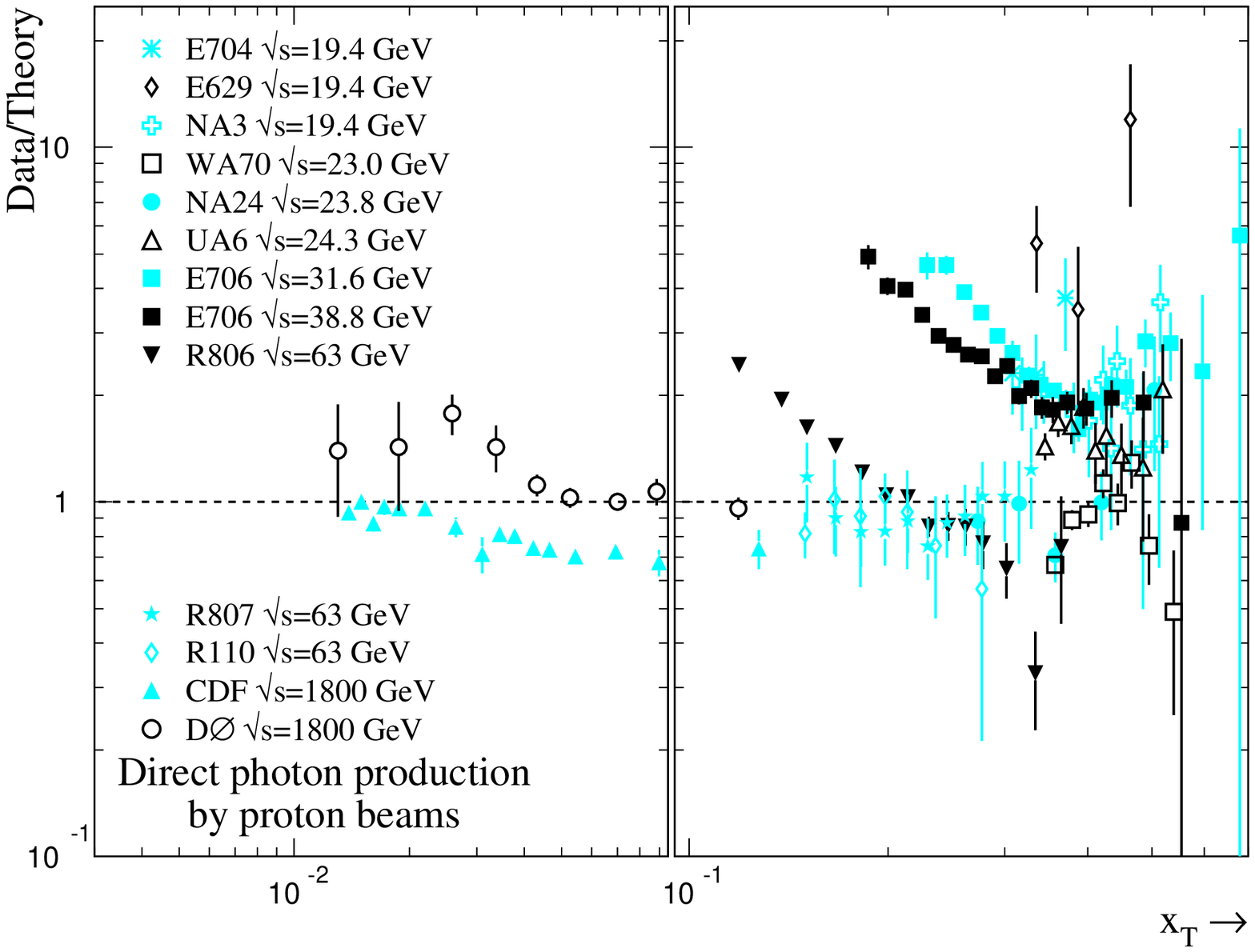}}
\caption{Comparison between proton-induced direct-photon data and NLO pQCD
calculations as a function of photon $x_T$.
\label{fig:dpdiscrepancy}}
\end{figure}

\begin{figure}
\epsfxsize=3truein
\centerline{\epsffile{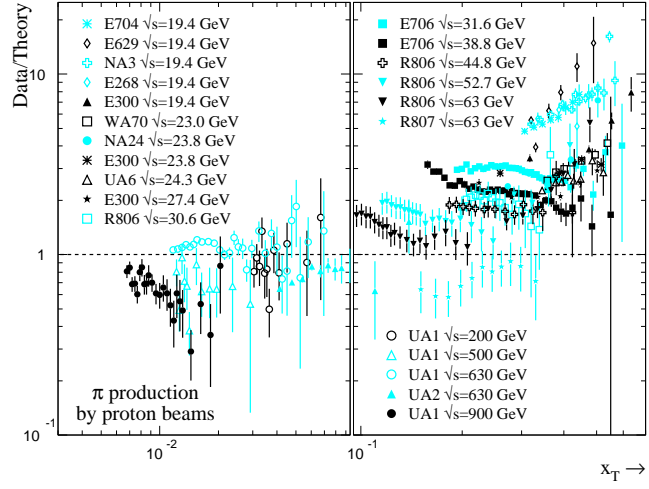}}
\caption{Comparison between proton-induced pion data and NLO pQCD
calculations for experiments spanning $\sqrt{s}=19.4$ to~$900$~GeV as
a function of $x_T$.
\label{fig:worldpi}}
\end{figure}

However, both the completeness of the theoretical NLO description of
the direct-photon process, as well as the consistency of results from
different experiments, have been subjects of intense
debate~\cite{mrst,baerreno,huston-discrepancy,ktprd,aurenche-dp,aurenche-pi0,kimber}.
The inclusive production of hadrons provides a further means of
testing the predictions of the NLO pQCD formalism.  Deviations have
been observed between measured inclusive direct-photon and pion cross
sections and NLO pQCD calculations.  Examples of such discrepancies
are shown in Figs.~\ref{fig:dpdiscrepancy} and~\ref{fig:worldpi},
where ratios of data to theory are displayed as a function of
$x_T=2p_T/\sqrt{s}$ for $p_T\agt3$~GeV/$c$ for photon and pion data,
respectively~\cite{E704,E629,NA3,WA70-p-dp,NA24,UA6,E706-kt,R806-dp-gpi,R806-pi0,R807-dp-pi0,R110-dp-gpi,compiled}.
(Unless otherwise indicated, all NLO
calculations~\cite{aurenche-nlo,aurenche-other,baer,berger,qiu,gordon,aversa}
in this paper use a single scale of $\mu=p_T/2$, CTEQ4M
PDFs~\cite{cteq4}, and BKK fragmentation functions for
pions~\cite{BKK}.)  The uncertainties shown in the figures reflect
mainly the statistical fluctuations in the measurements, except for
some pioneering experiments from the ISR, which reported statistical
errors combined with some of their systematic uncertainties.

Several of the older experiments quote large (of order 40\%)
uncertainties in their normalizations, making it difficult to judge
the significance of the discrepancies observed in
Figs.~\ref{fig:dpdiscrepancy} and~\ref{fig:worldpi}. Nevertheless, it
is clear that both the photon and pion data show disagreement with
theory, and the different data sets also disagree.  This is not
surprising, because, especially for direct-photon production, signals
are often very small and difficult to extract, and backgrounds are
quite large, especially at lower energies. Several experiments show
better agreement with NLO pQCD than others, but the results do not
provide great confidence in the theory nor in the quality of all the
data.  While it has been suggested that deviations from theory for
both photons and pions can be ascribed to higher-order effects of
initial-state soft-gluon radiation~\cite{huston-discrepancy,ktprd}, it
seems unlikely that theoretical developments alone will be able to
accommodate the level of scatter observed in
Figs.~\ref{fig:dpdiscrepancy} and~\ref{fig:worldpi}.

\section{$\gamma/\pi^0$ Ratio}

Given the scatter of the data shown in Figs.~\ref{fig:dpdiscrepancy}
and~\ref{fig:worldpi}, it may be instructive to consider measurements
of the $\gamma/\pi^0$ ratio in different experiments over a wide range
of $\sqrt{s}$.  Both experimental and theoretical uncertainties tend
to cancel in such a ratio, and this ratio should also be less
sensitive to incomplete treatment of gluon radiation. In addition,
since $\pi^0$ production constitutes the primary background to photon
production, the $\gamma/\pi^0$ ratio can serve as a measure of the
difficulty of the experimental environment for making the
direct-photon measurement.

\begin{figure}
\epsfxsize=3truein
\centerline{\epsffile{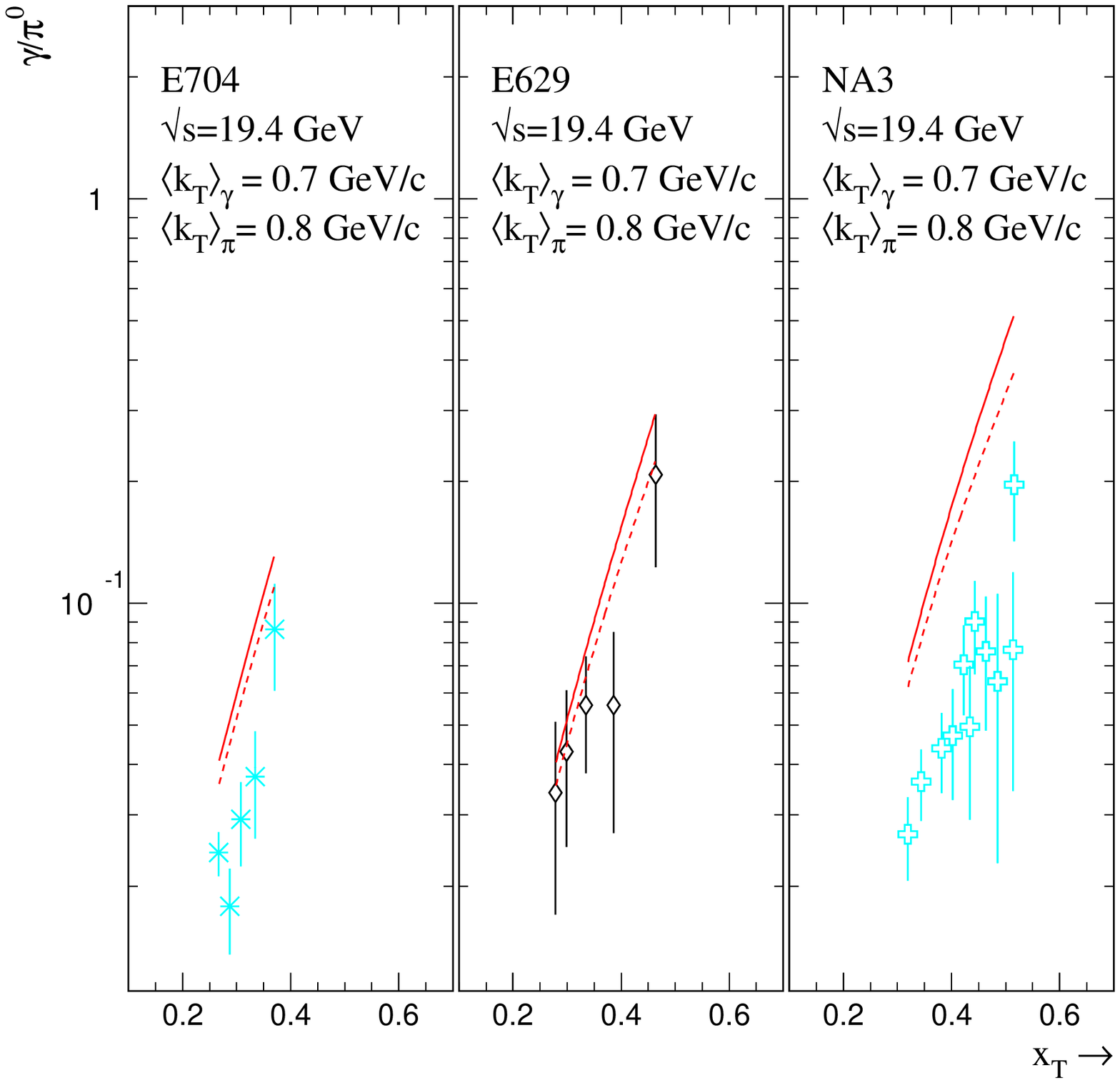}}
\caption{Comparison of $\gamma/\pi^0$ rates 
as a function of $x_T$ for E704, E629, and NA3 at $\sqrt{s}=19.4$~GeV.
Overlayed are the results from NLO pQCD (solid) and $k_T$-enhanced
calculations (dashed).  Values of $\langle k_T\rangle$ used for the
$k_T$-enhanced calculations are given in the legend.
\label{fig:photonTOpi0_1}}
\end{figure}

\begin{figure}
\epsfxsize=3truein
\centerline{\epsffile{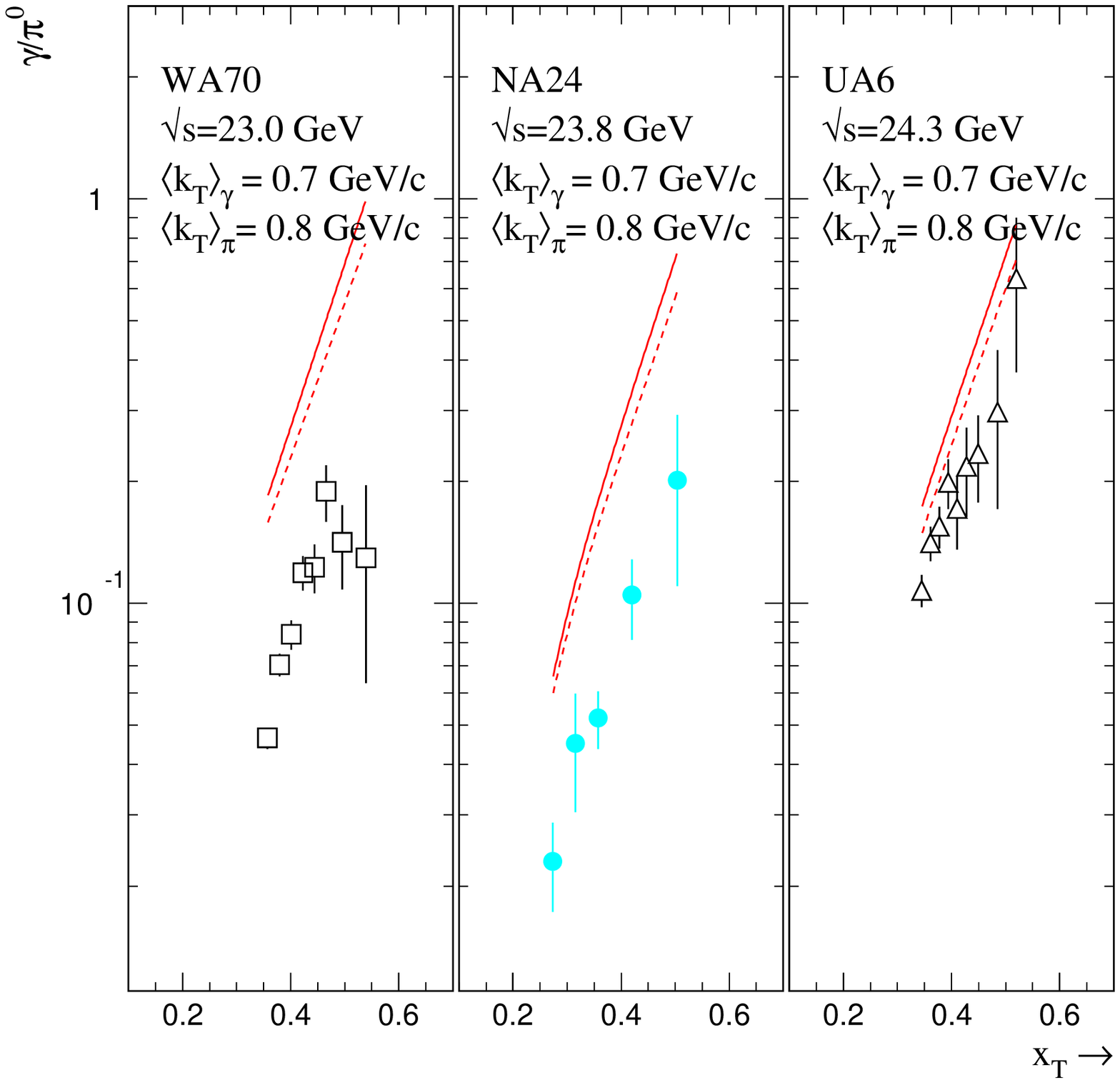}}
\caption{Comparison of $\gamma/\pi^0$ rates 
as a function of $x_T$ for WA70, NA24, and UA6 at
$\sqrt{s}\approx23-24$~GeV.  Overlayed are the results from NLO pQCD
(solid) and $k_T$-enhanced calculations (dashed).  Values of $\langle
k_T\rangle$ used for the $k_T$-enhanced calculations are given in the
legend.
\label{fig:photonTOpi0_2}}
\end{figure}

\begin{figure}
\epsfxsize=3truein
\centerline{\epsffile{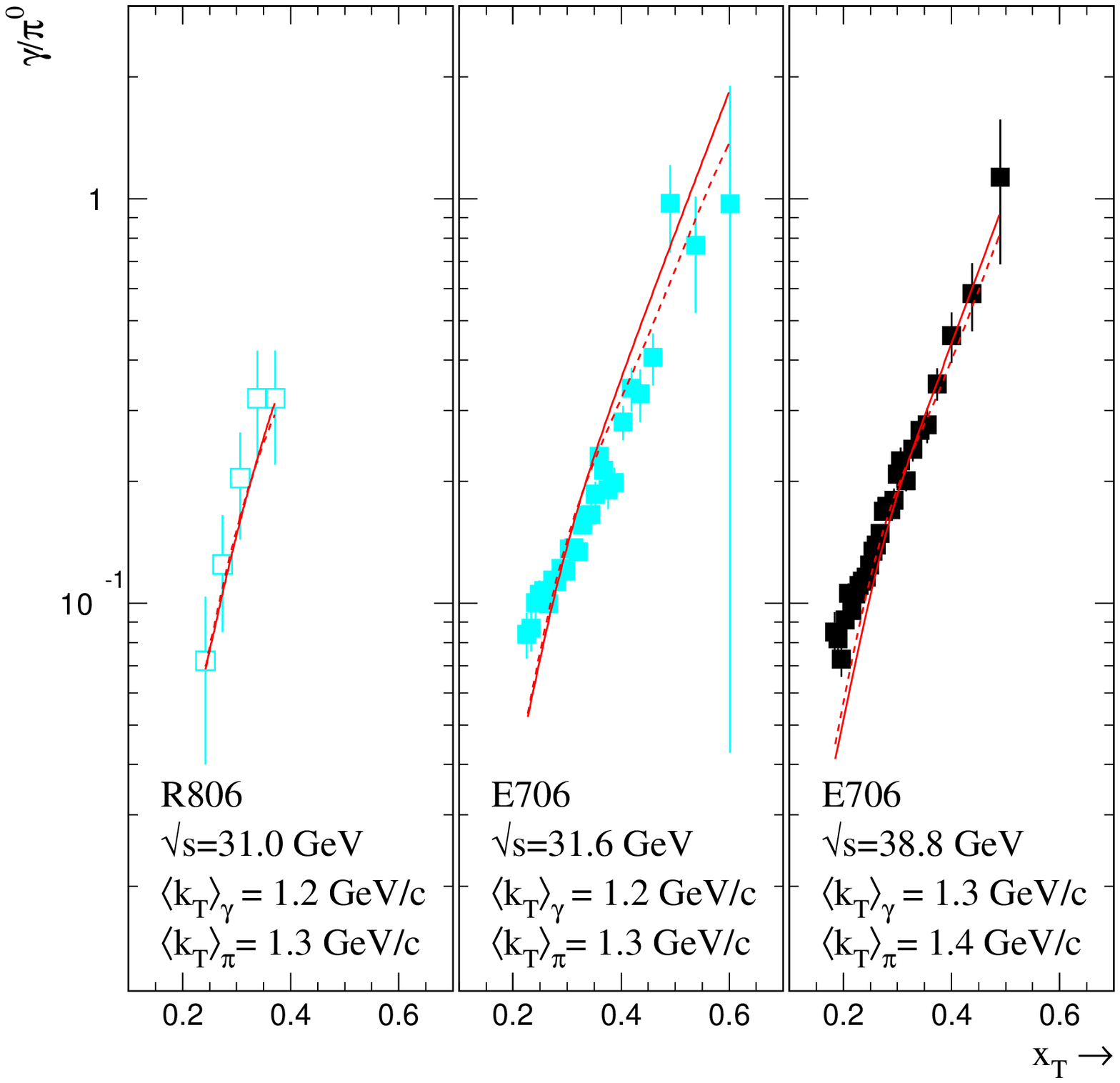}}
\caption{Comparison of $\gamma/\pi^0$ rates as a function of $x_T$ for 
R806 at $\sqrt{s}=31$~GeV and E706 at $\sqrt{s}=31.6$ and $38.8$~GeV.
Overlayed are the results from NLO pQCD (solid) and $k_T$-enhanced
calculations (dashed).  Values of $\langle k_T\rangle$ used for the
$k_T$-enhanced calculations are given in the legend.
\label{fig:photonTOpi0_3}}
\end{figure}

\begin{figure}
\epsfxsize=3truein
\centerline{\epsffile{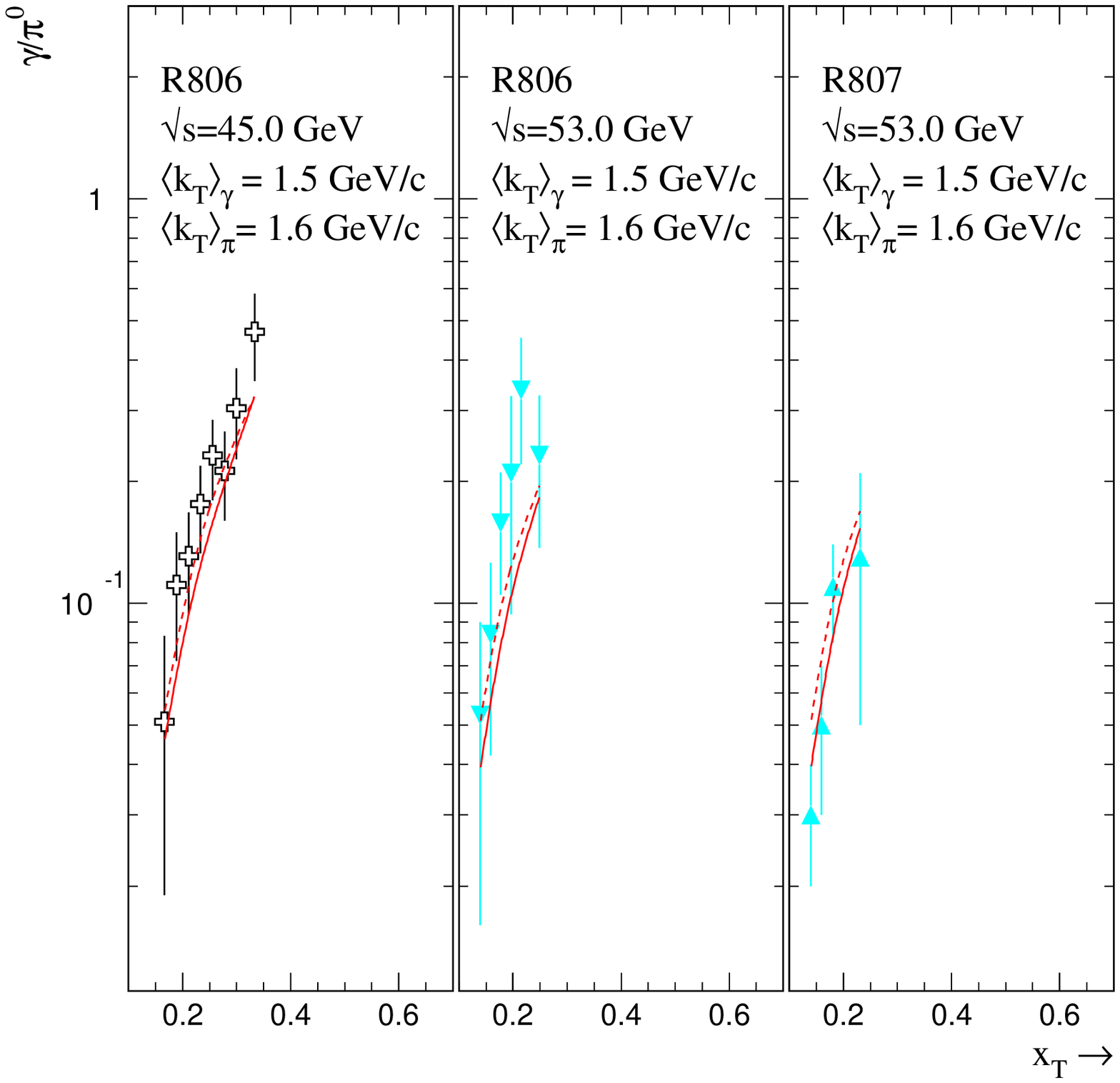}}
\caption{Comparison of $\gamma/\pi^0$ rates as a function of $x_T$ for 
R806 at $\sqrt{s}=45$ and $53$~GeV and R807 at $\sqrt{s}=53$~GeV.
Overlayed are the results from NLO pQCD (solid) and $k_T$-enhanced
calculations (dashed).  Values of $\langle k_T\rangle$ used for the
$k_T$-enhanced calculations are given in the legend.
\label{fig:photonTOpi0_4}}
\end{figure}

\begin{figure}
\epsfxsize=3truein
\centerline{\epsffile{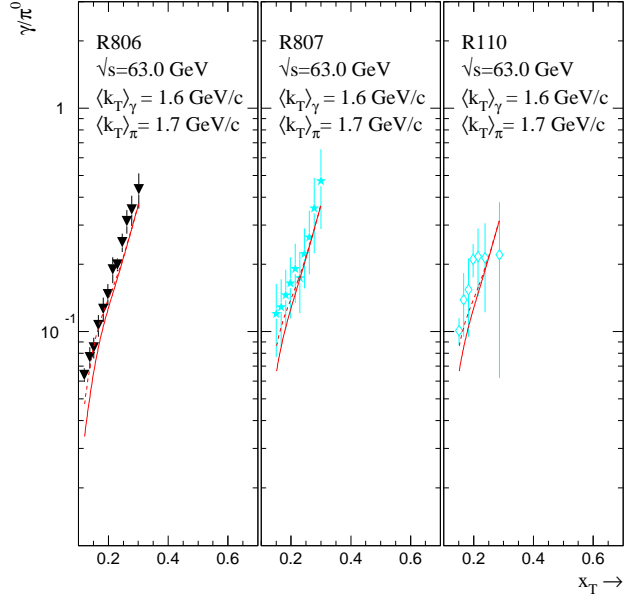}}
\caption{Comparison of $\gamma/\pi^0$ rates as a function of $x_T$ for 
R806, R807, and R110 at $\sqrt{s}=63$~GeV.  Overlayed are the results
from NLO pQCD (solid) and $k_T$-enhanced calculations (dashed).
Values of $\langle k_T\rangle$ used for the $k_T$-enhanced
calculations are given in the legend.
\label{fig:photonTOpi0_5}}
\end{figure}

The ratio of direct-photon to $\pi^0$ cross sections for both data and
NLO calculations (solid curves) are shown, as a function of $x_T$, in
Figs.~\ref{fig:photonTOpi0_1}--\ref{fig:photonTOpi0_5}.  Calculations
for experiments using nuclear targets (E629, NA3, NA24, E706) were
corrected approximately for nuclear dependence effects; these
corrections were $\alt 20$\% for $\pi^0$'s and $\alt10$\% for photons.
The results from E704~\cite{E704}, E629~\cite{E629}, and
NA3~\cite{NA3}, all at $\sqrt{s}=19.4$~GeV, are displayed in
Fig.~\ref{fig:photonTOpi0_1}.  For all three measurements, theory is
high compared to data, somewhat less so for E629.
Figure~\ref{fig:photonTOpi0_2} shows the~$\gamma$ to~$\pi^0$ ratio for
WA70~\cite{WA70-p-dp}, NA24~\cite{NA24}, and UA6~\cite{UA6} at
$\sqrt{s}\approx23-24$~GeV.  Just as for the comparisons at
$\sqrt{s}=19.4$~GeV, theory is high relative to data, with the
deviation between theory and data for WA70 and NA24 being greater than
that for UA6.

\begin{figure}
\epsfxsize=3truein
\centerline{\epsffile{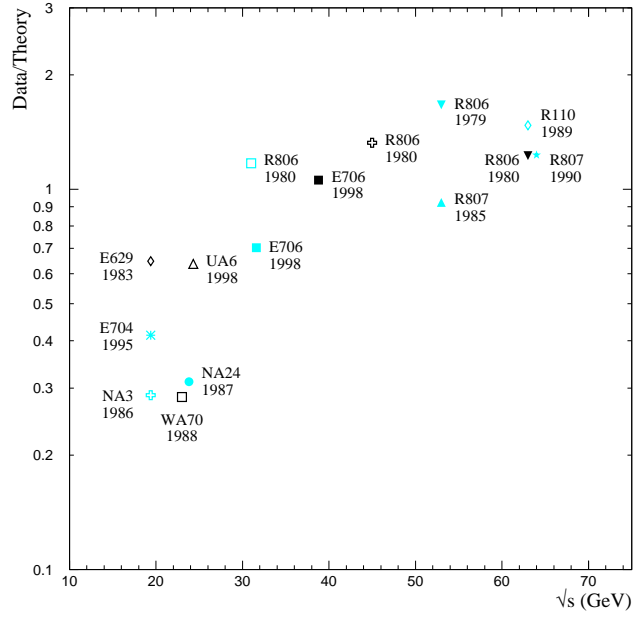}}
\caption{The ratio of data to theory for the $\gamma/\pi^0$ comparisons 
presented as a function of $\sqrt{s}$ for the direct-photon
experiments considered in this paper.  The values represent fits to
the ratio of data to NLO pQCD theory, without $k_T$-enhancement (see
text).
\label{fig:photonTOpi0_all}}
\end{figure}

For larger values of $\sqrt{s}$, the theoretically calculated value of
the $\gamma/\pi^0$ ratio agrees better with
experiment. Figure~\ref{fig:photonTOpi0_3} shows the data from
R806~\cite{R806-dp-gpi} at $\sqrt{s}=31$~GeV, and E706~\cite{E706-kt}
at $\sqrt{s}=31.6$ and $38.8$~GeV.  Theory provides a reasonable
description of these measurements. For the results shown in
Fig.~\ref{fig:photonTOpi0_4}, theory is now somewhat below the data
for R806~\cite{R806-dp-gpi} at $\sqrt{s}=45$ and~$53$~GeV, but agrees
with the R807 data~\cite{R807-gpi-53} at $\sqrt{s}=53$~GeV.  Theory is
also slightly low relative to data at $\sqrt{s}=63$~GeV for
R806~\cite{R806-dp-gpi,R806-pi0}, R807~\cite{R807-dp-pi0}, and
R110~\cite{R110-dp-gpi}, as shown in Fig.~\ref{fig:photonTOpi0_5}.  A
compilation of these results, shown for simplicity without their
uncertainties, is presented in Fig.~\ref{fig:photonTOpi0_all}.  In
this figure, the ratios of data to theory for the $\gamma$ to $\pi^0$
measurements shown in
Figs.~\ref{fig:photonTOpi0_1}--\ref{fig:photonTOpi0_5} have been
fitted to a constant value at high-$p_T$, and the results plotted as a
function of $\sqrt{s}$.  (Because the ratio depends somewhat on $p_T$,
we excluded from the fit points at low $p_T$, until a
$\chi^2/d.o.f.=1.5$ was reached for each fit.)  The figure suggests an
energy dependence in the ratio of data to theory for $\gamma/\pi^0$
production.  There is, however, an indication of substantial
differences between the experiments at low $\sqrt{s}$, where the
observed $\gamma/\pi^0$ is smallest, which makes it difficult to
quantify this trend.  Recognizing the presence of these differences is
especially important because thus far only the low energy photon
experiments have been used in PDF fits to extract the gluon
distribution.

\section{Phenomenological Corrections for Soft Gluon Emission}

In hadronic hard-scattering processes, there is generally a
substantial amount of effective parton transverse momentum, $k_T$, in
the initial state resulting from gluon emission~\cite{ktprd,begel}.
The presence of $k_T$ impacts the final state and has been observed in
measurements of Drell-Yan, diphoton, and heavy quark production; the
amount of $k_T$ expected from NLO calculations is not sufficient to
describe the data.  The effective values of $\langle
k_T\rangle$/parton for these processes vary from $\approx 1$~GeV/$c$
at fixed target energies to 3-4 GeV/$c$ at the Tevatron Collider. The
growth is approximately logarithmic with center-of-mass
energy~\cite{ktprd,begel}.  The size of the $\langle k_T\rangle$
values, and their dependence on energy, argue against a purely
``intrinsic'' non-perturbative origin.  Rather, the major part of this
effect is generally attributed to soft gluon emission.  While the
importance of including gluon emission through the resummation
formalism has long been recognized and such calculations have been
available for some time for Drell-Yan~\cite{altarelli},
diphoton~\cite{RESBOS,fergani}, and W/Z production~\cite{RESBOS}, they
have only recently been developed for inclusive direct-photon
production~\cite{nason,kidonakisowens,laenen,lilai,li,sterman}.

In the absence of a more rigorous theoretical treatment of the impact
of soft-gluon emission on high-$p_T$ inclusive production, a more
intuitive, but successful, phenomenological approach has been utilized
for describing the effects of soft-gluon radiation~\cite{ktprd}.  The
soft-gluon radiation was parametrized in terms of an effective
$\langle k_T\rangle$ that provided an additional transverse impulse to
the outgoing partons. Because of the steeply falling cross section in
$p_T$, such a $\langle k_T\rangle$ can shift the production of
final-state particles from lower to higher values of $p_T$,
effectively enhancing the cross section.  Using this intuitive
picture, the effects of soft gluon emission are approximated through a
convolution of the NLO cross section with a Gaussian $k_T$ smearing
function.  The value of $\langle k_T\rangle$ used for each kinematic
regime is based on experimental data~\cite{ktprd}.

\begin{figure}
\epsfxsize=3truein
\centerline{\epsffile{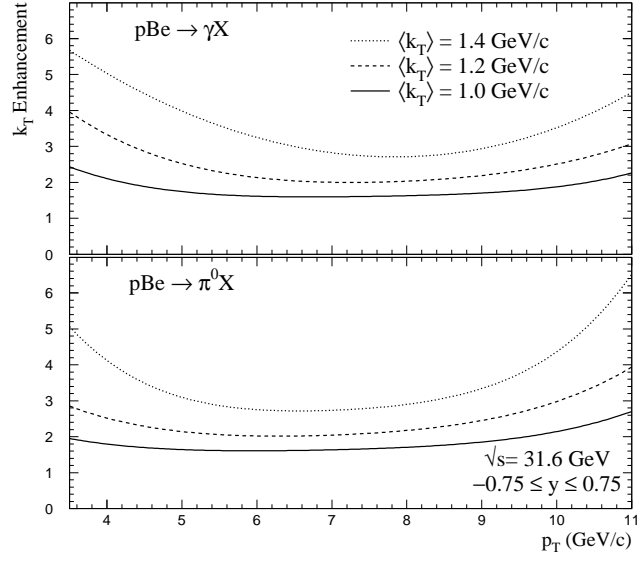}}
\caption{The dependence of the $k_T$-enhancement on $p_T$, $K(p_T)$, 
for variations in $\langle k_T\rangle$ for the E706 pBe data at
$\sqrt{s}=31.6$~GeV.
\label{fig:kfactors}}
\end{figure}

As described in Ref.~\cite{ktprd}, a leading order (LO) pQCD
calculation~\cite{owens} is used to generate correction factors
(ratios of calculations for any given $\langle k_T\rangle$ to the
result for $\langle k_T\rangle =0$) for inclusive cross sections
(Fig.~\ref{fig:kfactors}).  These $p_T$-dependent K-factors are then
applied to the NLO pQCD calculations.  This procedure involves a risk
of double-counting since some of the $k_T$-enhancements may already be
contained in the NLO calculation.  However, the effects of such
double-counting are expected to be small~\cite{ktprd}.  The
relationship between this phenomenological $k_T$-smearing and the
Collins--Soper--Sterman (CSS) resummation formalism~\cite{CSS} has
been discussed in Ref.~\cite{csaba-thesis}.

As illustrated in the upper part of Fig.~\ref{fig:kfactors}, the
K-factors for direct-photon production in E706 are large over the full
range of $p_T$, and have $p_T$-dependent shapes. The lower part of
Fig.~\ref{fig:kfactors} displays K-factors for $\pi^0$ production,
based on the same model. For the same values of $\langle k_T\rangle$,
the K-factors in $\pi^0$ production are somewhat smaller than in
$\gamma$ production.  This is reasonable because $\pi^0$ mesons
originate from the fragmentation of partons, and therefore carry only
a fraction of the $k_T$ given to the jet.  Because the momentum
fraction of the jet carried by a $\pi^0$ tends to increase with
increasing $p_T$ of the $\pi^0$ (sometimes referred to as
trigger-bias), the $k_T$ corrections in the case of $\pi^0$ production
are smaller and tend to have a weaker $p_T$-dependence than the
corrections for direct photons.

In this model, the size of the correction depends sensitively on the
value used for $\langle k_T\rangle$. Changes of 200 MeV/$c$ produce
substantial differences in the size of the correction.  Values of
$\langle k_T\rangle$ are only accurate to about this
range~\cite{ktprd}, and so, even within the context of this model, it
is difficult to obtain the kind of precision needed for extracting
global parton distributions.  In addition, there are different ways to
implement the $k_T$-corrections~\cite{mrst}, which can produce
quantitative differences in the magnitude and shape of the $k_T$
correction factors.

\begin{figure}
\epsfxsize=3truein
\centerline{\epsffile{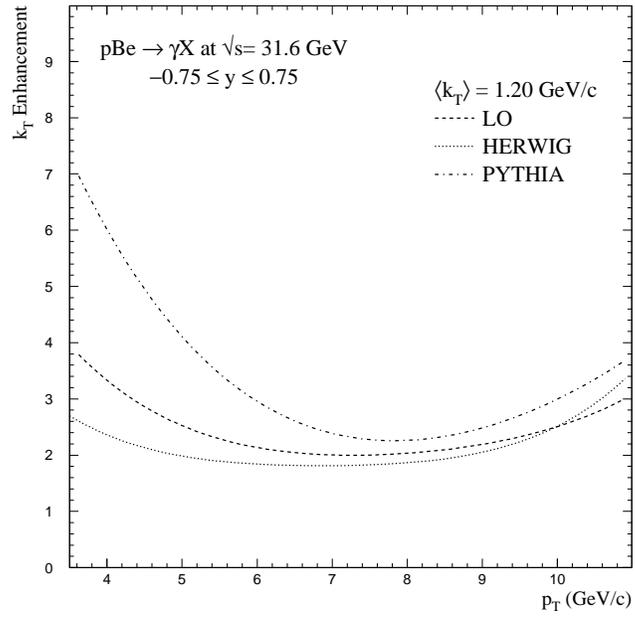}}
\caption{Comparison between the $k_T$ enhancements, $K(p_T)$, from
the LO pQCD calculation (dashed), {\sc herwig} (dotted), and {\sc
pythia} (dashed-dotted).
\label{fig:kfactors_mc}}
\end{figure}

\begin{figure}
\epsfxsize=3truein
\centerline{\epsffile{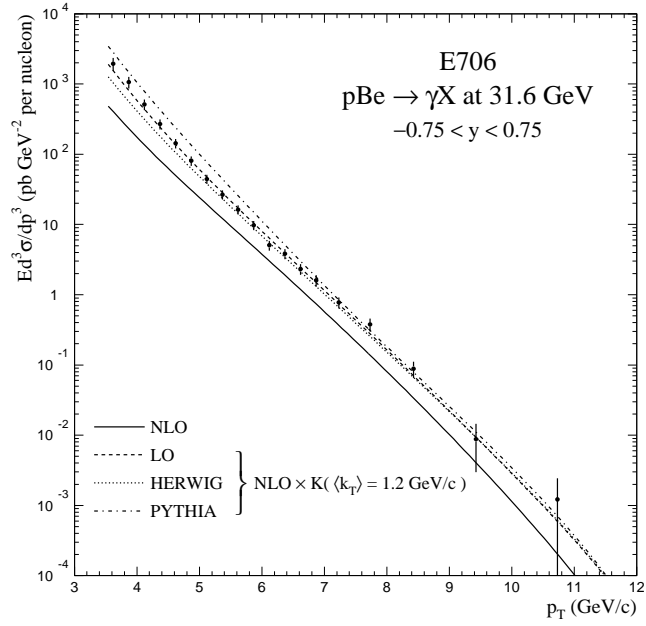}}
\caption{Comparison between the E706 direct-photon data at 
$\sqrt{s}=31.6$~GeV and the NLO pQCD calculation(solid), and the NLO
theory enhanced by K-factors associated with the LO calculation
(dashed), {\sc herwig} (dotted), and {\sc pythia} (dashed-dotted).
\label{fig:kfactors_e706}}
\end{figure}

We have investigated the kind of K-factors that would be expected for
direct-photon production from parton-showering
models~\cite{pythia61,herwig61}. These programs do not provide
sufficient smearing at fixed-target energies because shower
development is constrained by cut-off parameters that ensure the
perturbative nature of the process.  Consequently, these calculations
allow additional input $k_T$ for Gaussian smearing, and are often used
that way in comparisons to data (default values are $\langle
k_T\rangle=0$ for {\sc herwig} and $\langle k_T\rangle=0.9$~GeV/$c$
for {\sc pythia}).  Using default settings for other program
parameters and an input $\langle k_T\rangle$ of 1.2 GeV/$c$ (to
compare to our previous results~\cite{ktprd}) for the smearing,
relative to these same settings with $k_T=0$ in these generators, we
obtain the K-factors shown in Fig.~\ref{fig:kfactors_mc}. These
corrections are different than those found using the LO pQCD
prescription~\cite{owens}.  These differences must be kept in mind
when making comparisons to data.  Applying the K-factors in
Fig.~\ref{fig:kfactors_mc} to NLO pQCD results in the comparison to
data from E706 shown in Fig.~\ref{fig:kfactors_e706}.

The dashed curves in
Figs.~\ref{fig:photonTOpi0_1}--\ref{fig:photonTOpi0_5} describe the
predicted ratios of $\gamma$ and $\pi^0$ cross sections using the
previous $k_T$ corrections~\cite{ktprd}.  The impact of $k_T$
corrections upon $\gamma/\pi^0$ is generally minimal, indicating that
the trend observed in Fig.~\ref{fig:photonTOpi0_all} cannot be easily
understood on the basis of corrections for $k_T$ alone.

\section{Recent Progress in Theory}

Resummed pQCD calculations for single direct-photon production are
currently under
development~\cite{nason,kidonakisowens,laenen,lilai,li,sterman}.
Substantial corrections to fixed-order QCD calculations are expected
from soft-gluon emission, especially in regions of phase space where
gluon emission is restricted kinematically.  Such restrictions occur,
for example, in the production of Drell-Yan or diphoton pairs at low
$p_T$, where effects of soft-gluon emission have to be resummed in
order to obtain adequate description of cross
sections~\cite{altarelli}.  In addition, at high $x$, there is a
similar suppression of gluon radiation due to the rapidly falling
parton distributions. A complete description of the cross section in
this region requires the resummation of ``threshold'' terms.  For
inclusive ``single-arm'' observables, there is no restriction in phase
space for soft gluon radiation, and therefore no enhancement to the
cross section is expected from traditional resummation calculations
except at high $x$~\cite{LHCworkshopdocument}.  Two recent independent
threshold-resummed pQCD calculations for direct
photons~\cite{nason,kidonakisowens} exhibit far less dependence on QCD
scales than found in NLO theory.  These calculations agree with the
NLO prediction for the scale $\mu\approx p_T/2$ at low $p_T$ (without
inclusion of explicit $k_T$ or recoil effects), and show an
enhancement in cross section at high~$p_T$ due to the threshold terms.

A method for simultaneous treatment of recoil and threshold
corrections in inclusive single-photon cross sections is being
developed~\cite{sterman} within the formalism of collinear
factorization.  This approach accounts explicitly for the recoil from
soft radiation in the hard-scattering subprocess, and conserves both
energy and transverse momentum for the resummed radiation.  The
possibility of substantial enhancements from higher-order perturbative
and power-law nonperturbative corrections relative to NLO are
indicated at both moderate and high $p_T$ for fixed-target energies,
similar to the enhancements obtained with the simple $k_T$-smearing
model discussed above.

\begin{figure}
\centering\leavevmode
\vglue1truept\vspace{-3cm}
\hglue1truept\hspace{14cm}
\epsfxsize 9 cm 
\epsfbox{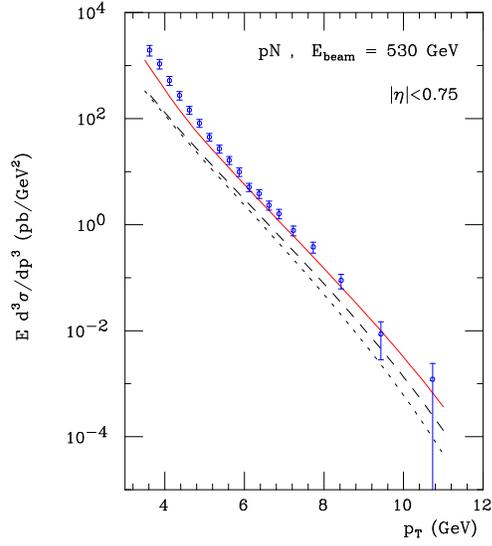}
\vglue1truept\vspace{1.5cm}
\caption{Direct-photon cross section for the E706 data at 
$\sqrt{s}=31.6$~GeV compared to recent QCD calculations.  The dotted
line represents the full NLO calculation~\protect\cite{gordon}, while
the dashed and solid lines, respectively, incorporate purely threshold
resummation~\protect\cite{nason} and joint threshold and recoil
resummation~\protect\cite{sterman}.
\label{fig:sterman}} 
\end{figure}

Figure~\ref{fig:sterman} displays the results of an example
calculation based on the latest approach involving both threshold and
recoil resummation for comparison with direct-photon measurements from
E706~\cite{sterman}.  As shown in Fig.~\ref{fig:sterman}, this
theoretical result is substantially higher than the prediction from
NLO pQCD, higher than the theory using just threshold resummation, and
closer to the phenomenological $k_T$-smearing model used in
Fig.~\ref{fig:kfactors_e706}.

\section{Conclusions}

We have examined experimental information on the production of direct
photons and pions at large $p_T$ and reviewed how LO pQCD has been
used to estimate the impact of $k_T$ on their inclusive production.
Simple phenomenological models can improve agreement between pQCD
calculations and much of the inclusive data over a wide range
of~$\sqrt{s}$~\cite{ktprd}.  While the deficiencies of such models are
clear, and have been discussed in recent literature, the emerging
formalism for the full (threshold and recoil) resummation of inclusive
direct-photon cross sections~\cite{sterman} appears to support some of
the understanding of soft multiple gluon emission that has been gained
using the approximate $k_T$ tools.

Despite the apparent inconsistencies between different direct-photon
experiments~\cite{ktprd,aurenche-dp,aurenche-pi0}, we found it
instructive to consider results on the measured $\gamma/\pi^0$ ratios
since various experimental and theoretical uncertainties tend to
cancel in such ratios.  We find that the ratios taken from theory
agree reasonably with those computed from data for
$\sqrt{s}\agt30$~GeV. However, there appears to be a systematic trend
with energy that is unexplained, and differences between experiments
that are particularly signficant at lower energy.

While there is still no resummation calculation for inclusive pion
production, the trend of recent developments in direct-photon
processes has led to an increased appreciation of the importance of
the effect of multiple gluon emission, and to the emergence of tools
for incorporating these effects.  These latest theoretical
developments encourage optimism that the long-standing difficulties in
developing an adequate description of these processes can eventually
be resolved, making possible a global re-examination of parton
distributions with an emphasis on the determination of the gluon
distribution from the direct-photon data~\cite{sterman-pdf}.

\acknowledgments

We wish to thank P.~Aurenche, C.~Bal\'{a}zs, S.~Catani, M.~Fontannaz,
N.~Kidonakis, J.~F.~Owens, E.~Pilon, G.~Sterman, and W.~Vogelsang for
helpful discussions.

\renewcommand{\baselinestretch}{1.}
\newpage
\bibliography{paper}  
\bibliographystyle{prsty}
%
%
\widetext
\newpage

\newpage

\newpage

\newpage

\newpage

\newpage

\newpage

\newpage

\newpage

\newpage

\newpage

\newpage

\end{document}